\documentclass[a4paper,10pt]{article}

\usepackage{verbatim}
\usepackage{amsfonts}
\usepackage{amsmath}
\usepackage{mathrsfs}
\usepackage{amssymb}
\usepackage{amsthm}
\usepackage{graphicx}
\usepackage{psfrag}
\usepackage[bookmarks,colorlinks=false]{hyperref}

\newcounter{mnotecount}

\newcommand{\mnotex}[1]
{\protect{\stepcounter{mnotecount}}$^{\mbox{\footnotesize$
\bullet$\themnotecount}}$ \marginpar{
\raggedright\tiny\em
$\!\!\!\!\!\!\,\bullet$\themnotecount: #1} }

\newtheorem{theorem}{Theorem}
\newtheorem{prop}{Proposition}

\newtheorem{corollary}{Corollary}

\begin{document}

\title{A note on time-symmetric hypersurfaces in the Schwarzschild geometry}

\author{Alfonso Garc\'{i}a-Parrado G\'{o}mez-Lobo 
\thanks{E-mail adress: {\tt alfonso@math.uminho.pt}}\\
Centro de Matem\'atica, Universidade do Minho\\ 
4710-057 Braga, Portugal}

\maketitle


\begin{abstract}
In this note we show that any inextensible time-symmetric space-like hypersurface of differentiability class $C^k$, $k\geq 2$ isometrically embedded in the maximal Schwarzschild geometry must intersect the bifurcation sphere. 
\end{abstract}


\section{Introduction}
The Schwarzschild solution is one of the most relevant vacuum solution of the Einstein field equations and its geometric properties have been widely studied. Among these we quote the analysis of space-like embedded hypersurfaces (submanifolds of co-dimension 1) 
fulfilling certain geometric requirements.
These requirements have been mostly inspired by the analysis {\em the constraint equations} which arise from the orthogonal splitting of the Einstein field equations. These equations adopt a simpler form if one makes assumptions on the extrinsic curvature and it is therefore necessary to ensure that the corresponding assumptions can be made. The most common assumptions are to impose that the extrinsic curvature vanishes or has zero trace and this naturally leads to the question about the existence of embedded hypersurfaces possessing these properties. When the extrinsic curvature vanishes then the hypersurface is {\em totally geodesic}. The set of points which are invariant under a {\em time-reflexion} isometry with respect to a given time function forms a space-like hypersurface with vanishing extrinsic curvature and for that reason one uses the nomenclature ``time-symmetric'' for these hypersurfaces.

In \cite{REINHART} all the foliations of the Schwarzschild-Kruskal space-time whose leaves are spherically symmetric maximal hypersurfaces were found. This work was later generalised to the case in which the 
extrinsic curvature trace is a constant \cite{BRILL-ISENBERG}. Examples of these foliations can be found in \cite{SIDDIQUI-K,PERVEZ-SIDDIQUI}.
A maximal slicing of a subset of the Schwarzschild-Kruskal space-time containing a Cauchy hypersurface was constructed in 
\cite{BEIG-NIAL} and the implications in numerical simulations were addressed. A different kind of hypersurfaces were considered in 
\cite{PAREJA-FRAUENDIENER} where spherically symmetric hypersurfaces 
having constant scalar curvature were constructed. Yet another kind of foliation is that composed by hypersurfaces whose leaves have vanishing Riemann tensor (flat foliations). The uniqueness of 
these foliations in static spherically symmetric spacetimes was addressed in \cite{BEIG-SIDDIQUI}
and explicit examples were constructed in the Schwarzschild and Reissner Nordstr\"om solutions, \cite{SIDDIQUI-SCHWARZSHILD} and in the Kottler-Schwarzschild-de Sitter spacetime,  \cite{SIDDIQUI-RNKOTTLER}. 

In this note we study embedded, time-symmetric hypersurfaces in the maximal extended Schwarzschild-Kruskal space-time. The most trivial example of such a hypersurface is comprised by any integral hypersurface of the static Killing  
vector but no other instance of such a hypersurface is known. A natural question is whether these hypersurfaces are indeed the only embedded time-symmetric hypersurfaces which exist in the Schwarzschild-Kruskal space-time. We believe that the answer to this question is affirmative but so far a complete proof of this issue is lacking. However, one can show other interesting properties of embedded time-symmetric hypersurfaces in the Schwarzschild-Kruskal space-time. It is the aim of the present short note to show that any embedded time-symmetric hypersurface in the Schwarzschild-Kruskal space-time cannot intersect the {\em black hole region}. To the best of our knowledge, a rigorous proof of this statement is not available in the literature. 
A consequence of this result is that in the Schwarzschild geometry, all inextensible time-symmetric hypersurfaces embedded in the Schwarzschild-Kruskal space-time should intersect the bifurcation sphere (see corollary \ref{cor:hypersurface-no-boundary}).  

The outline of this work is as follows: in section \ref{sec:preliminaries} we summarise some known results about the Schwarzschild solution which are needed in this work. Section \ref{sec:main-results} contains the main results of this work which are the non-existence of embedded time-symmetric hypersurfaces in the non-static regions of the Schwarzschild space-time (proposition \ref{prop:time-symmetric}) and its corollary. Finally,
we discuss some applications in section \ref{sec:conclusions}.

\section{Preliminaries}
\label{sec:preliminaries}
\begin{figure}
\begin{center}
\psfrag{EHf}{${\mathcal H}^+$}
\psfrag{EHp}{${\mathcal H}^-$}
\psfrag{Jf}{${\mathscr J}^+$}
\psfrag{Jp}{${\mathscr J}^-$}
\psfrag{Om}{$\Omega$}
\psfrag{D1}{${\mathcal D}_1$}
\psfrag{D2}{${\mathcal D}_2$}
\psfrag{Bf}{${\mathcal B}^+$}
\psfrag{Bp}{${\mathcal B}^-$}
\includegraphics[width=.8\textwidth]{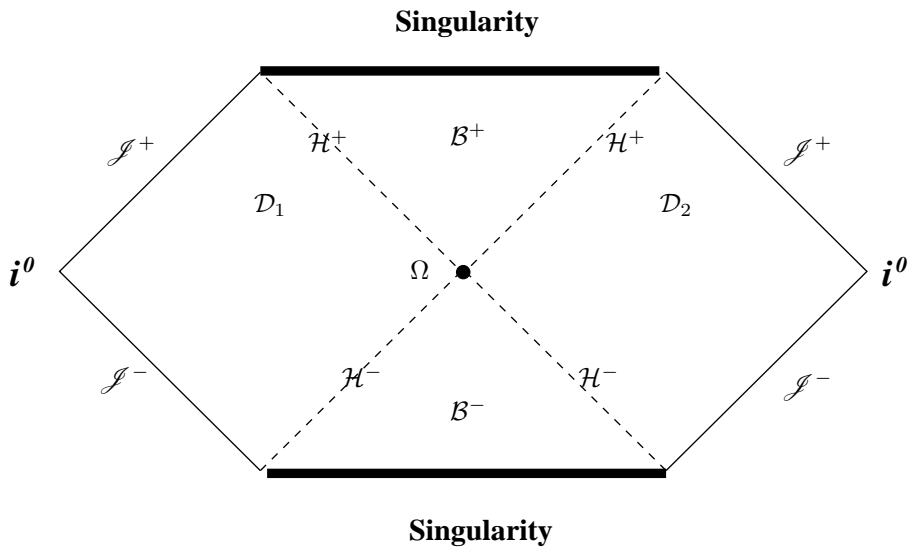}
\end{center}
\caption{\label{fig:penrose-kruskal} Penrose diagram of the Kruskal extension of the Schwarzschild space-time. We indicate the symbols used in the text for the different regions in order to fix 
the notation. These are the static 
regions ${\mathcal D}_1$, ${\mathcal D}_2$,
the future ${\mathcal B}^+$ and past  ${\mathcal B}^+$ black hole regions, 
the future ${\mathcal H}^+$, and past ${\mathcal H}^-$ event horizons, 
the bifurcation sphere $\Omega={\mathcal H}^+\cap {\mathcal H}^-$, 
the null infinities $\mathscr{J}^+$, $\mathscr{J}^-$ 
and space-like infinity $i^0$. 
We assume the reader familiar with the main features of these regions.}
\end{figure}

The maximal extension of the Schwarzschild geometry is the well-known Schwarzschild Kruskal space-time \cite{KRUSKAL} and this is what will be understood in this work by the Schwarzschild space-time. This is a globally hyperbolic Lorentzian manifold, which we shall denote by $\mathcal M$ with Lorentzian metric $g_{ab}$ (small Latin letters $a,b,c,\dots$ will be used to denote indices for tensors fields on any tensor bundle built from $T(\mathcal M)$ and its dual $T^*(\mathcal M)$). 
The global causal structure of the Kruskal maximal extension is well-known and it is displayed in figure \ref{fig:penrose-kruskal} where the different regions of this space-time are represented.  

Now let $\Sigma\subset{\mathcal M}$ be a $C^k$, $k>2$ embedded space-like submanifold and let $\vec{\boldsymbol N}$ be the unit timelike normal vector to $\Sigma$ (our signature convention is such that 
$g(\vec{\boldsymbol N},\vec{\boldsymbol N})=-1$). Unless otherwise stated, all hypersurfaces considered in this work will be space-like. 
The first fundamental form $h_{\alpha\beta}$ and the extrinsic curvature $K_{\alpha\beta}$ are defined by the standard formulae
\begin{equation}
h_{\alpha\beta}\equiv g(\vec{\boldsymbol e}_{\alpha},\vec{\boldsymbol e}_{\beta})\;,\quad
K_{\alpha\beta}\equiv g(\vec{\boldsymbol e}_{\beta},
\nabla_{\vec{\boldsymbol e}_{\alpha}}\vec{\boldsymbol N}),
\label{eq:first-second-form}
\end{equation}
where $\{\vec{\boldsymbol e}_{\alpha}\}$, $\alpha=1,2,3$ is an arbitrary frame on $\Sigma$ and $\nabla$ represents the Levi-Civita connection on $\mathcal M$. As is well-known the first fundamental form and the extrinsic curvature can be regarded as symmetric tensor fields on $\Sigma$ (indices for tensor fields on any tensor bundle with $\Sigma$ as the base manifold will be denoted with Greek characters $\alpha,\beta,\dots $). The pair $(\Sigma,h_{\alpha\beta})$ is a Riemannian manifold and therefore one can introduce the Levi Civita connection $D_\gamma$ of the Riemannian metric $h_{\alpha\beta}$. An embedded hypersurface $\Sigma$ is said to be {\em inextensible} if it is not a proper subset of another embedded hypersurface in $\Sigma$.
When the extrinsic curvature is zero on every point of $\Sigma$ then we say that it is a {\em time-symmetric hypersurface}. 

The Schwarzschild space-time can be characterised {\em locally} in terms of scalar and tensor {\em concomitants} of the Weyl tensor $W_{abcd}$ \cite{FERSAEZSCH}. Since some of the conditions arising from this characterisation are required later on, 
we review next its main aspects. First of all
let us define the tensor 
\begin{equation}
G{}_{a}{}_{b}{}_{c}{}_{d} \equiv g{}_{a}{}_{c} g{}_{b}{}_{d}-g{}_{a}{}_{d} g{}_{b}{}_{c}.
\label{eq:2-form-metric}
\end{equation}
The tensor $G{}_{a}{}_{b}{}_{c}{}_{d}$ has algebraic properties similar to those of the Riemann tensor. Next we introduce the Weyl tensor  {\em scalar concomitants}
\begin{equation}
\rho\equiv \frac{1}{2}
\left(\frac{W_{ab}^{\phantom{ab}cd}W_{cd}^{\phantom{cd}pq}W_{pq}^{\phantom{pq}ab}}{12} \right)^{1/3}\;,
\quad
\alpha\equiv 2\rho+\frac{\nabla_a\rho\nabla^a\rho}{9\rho^2},
\label{eq:weyl-scalars}
\end{equation}
and {\em tensorial concomitants}
\begin{equation}
S_{abcd}\equiv\frac{1}{3\rho}(W_{abcd}+\rho G_{abcd}),\quad
Q_{ab}\equiv S_{apbq}\nabla^p\rho \nabla^q\rho,\quad
P_{ab}\equiv W^*_{apbq}\nabla^p\rho \nabla^q\rho,
\end{equation}
where $W^*_{apbq}\equiv\eta_{bqcd}W_{ap}^{\phantom{ap}cd}/2$ and $\eta_{abcd}$ is the volume element 
associated to the metric $g_{ab}$.

One has now the following result proven in \cite{FERSAEZSCH}.
\begin{theorem}[\bf Ferrando \& S\'aez, (1998)]
A space-time $({\mathcal M}, g_{ab})$ is locally isometric to the Schwarzschild solution if and only if the following conditions are fulfilled
\begin{eqnarray}
&&- S{}_{a}{}_{b}{}_{p}{}_{q} + \frac{1}{2} S{}_{a}{}_{b}{}^{c}{}^{d}
S{}_{c}{}_{d}{}_{p}{}_{q} = 0\;,\quad
P_{ab}=0\;,\quad  
\alpha>0\;,
\nonumber\\
&&Q{}_{a}{}_{b} Q{}^{b}{}_{c}-Q{}_{a}{}_{c}
\nabla{}_{p}\rho\nabla{}^{p}\rho  = 0\;,\quad \rho\neq 0.
\label{eq:sch-conditions}
\end{eqnarray}
\label{theo:schwarzschild-charact}
\end{theorem}

\section{Main results}
\label{sec:main-results}
\begin{prop}
Let $\Sigma\subset{\mathcal M}$ be an embedded $C^k$, $k>2$ hypersurface 
and assume further that $\Sigma$ is time-symmetric. Then $\Sigma$ does not intersect the interior of the black hole region of 
the Kruskal maximal extension.
\label{prop:time-symmetric}
\end{prop}
\proof Define a one-parameter family of embedded $C^k$ space-like hypersurfaces $\{\Sigma_{t}\}_{t\in I}$, where $I$ is an open real interval, $\Sigma_{t_1}\cap\Sigma_{t_2}=\varnothing$ if $t_1\neq t_2$ and $\Sigma_0=\Sigma$. The family $\{\Sigma_{t}\}$ can then be regarded as a foliation of the set 
$\mathcal U\equiv \cup_{t\in I}\Sigma_t$. The unit timelike normal to each leave $\Sigma_{t}$ is a 
$C^k$ vector field on int$(\mathcal U)$ and we shall denote this vector field by $n^a$. We take now the conditions (\ref{eq:sch-conditions}) and compute their orthogonal splitting with respect to the vector field $n^a$. Such a computation was carried out in \cite{GARVALSCH} and we shall only quote here the result which we need, referring the reader to the afore-mentioned reference for a complete derivation. The needed result is the orthogonal splitting of the gradient of the scalar $\rho$
which reads 
\begin{equation}
\nabla_a\rho=P n_a+P_a\;,
\label{eq:decompose-rho}
\end{equation}
where
\begin{subequations}
\begin{eqnarray}
&&P\equiv -\frac{1}{2} {E}_{{  }}^{{ab }}
{K}_{{ab }}^{{  }}-{\rho}_{} {K}_{{\ a}}^{{a }}
-\frac{1}{6{\rho}_{}}\epsilon _{{\ \ \ a}}^{{d c}}
\left({E}_{{  }}^{{ab }} D_c B_{{b d}}^{{  }}
+B_{{  }}^{{ab }} D_c{E}_{{b d}}^{{  }}\right), \label{eq:scalarP}\\
&&P_c\equiv\frac{1}{6 {\rho}_{}}(-B_{{  }}^{{ab }}
D_c B_{{ab }}^{{  }}+{E}_{{  }}^{{ab }} D_c{E}_{{ab}}^{{  }})=D_c\rho,
\label{eq:vectorP}
\end{eqnarray}
\end{subequations}  
and 
\begin{eqnarray}
&& 
h_{ab}\equiv g_{ab}+n_{a}n_{b}\:,\quad 
K_{ab}\equiv-\frac{1}{2}\mathcal{L}_{\vec{\boldsymbol n}}h_{ab}\:,\quad
\epsilon_{abc}\equiv\eta_{d abc}n^{d},
\label{eq:lie-h}\\
&&E_{ab}\equiv W_{acbd}n^cn^d\;, \quad 
B_{ab}\equiv W^*_{acbd}n^cn^d.
\label{eq:weyl-e-m}
\end{eqnarray}
The operator $D_c$ appearing in some of the previous equations 
is defined for any tensor field $T^{a_1\dots a_p}_{\ b_1\dots b_q}$ by 
\begin{equation}
D_c T^{a_1\dots a_p}_{\ b_1\dots b_q}\equiv
h^{a_1}_{\phantom{a_1}d_1}\dots h^{a_p}_{\phantom{a_p}d_p}
h^{s_1}_{\phantom{s_1}b_1}\dots
h^{s_q}_{\phantom{s_q}b_q}h^{l}_{\phantom{l}c}\nabla_l
T^{d_1\dots d_p}_{\ s_1\dots s_q}.
\end{equation}
From (\ref{eq:decompose-rho})
we easily deduce 
\begin{equation}
\nabla_a\rho\nabla^a\rho=-P^2+P_a P^a. 
\label{eq:grad-rho-norm}
\end{equation}
We realise that $P$ and $P_c$ are given in terms of {\em spatial quantities} (quantities invariant under the action of the spatial projection $h_a^{\phantom{a}b}$) and that makes it easy to compute the pull-back of them to any of the hypersurfaces $\Sigma_t$ under the inclusion embedding
$i:\Sigma_t\subset\mathcal M$. 
In particular the pull-back of $h_{ab}$ ($K_{ab}$)
under this embedding is the first fundamental form (the extrinsic curvature) of $\Sigma_t$ which is denoted by 
$h^{(t)}_{\alpha\beta}$ ($K^{(t)}_{\alpha\beta}$). Also, the pull-backs of $E_{ab}$, $B_{ab}$ can be rendered
in the form (see again \cite{GARVALSCH})
\begin{equation}
E^{(t)}_{\alpha\beta}\equiv r^{(t)}_{\alpha\beta}+K^{(t)\gamma}_{\phantom{(t)\gamma}\gamma} K^{(t)}_{\alpha\beta}-K^{(t)}_{\alpha\gamma}K^{(t)\gamma}_{\phantom{\gamma}\beta}\;,\quad
B^{(t)}_{\alpha\beta}\equiv\epsilon_{(\alpha}^{(t)\phantom{\alpha}\gamma\delta}D^{(t)}_{|\gamma}
K^{(t)}_{\delta|\beta)},
\label{eq:electric-magnetic}
\end{equation}
where $r^{(t)}_{\alpha\beta}$ is the Ricci tensor computed from the curvature arising from the covariant derivative $D^{(t)}_{\alpha}$
(we follow the convention of adding the superscript $t$ to those quantities intrinsic to the 
Riemannian manifold $(\Sigma_t, h^{(t)}_{\alpha\beta})$). Now, by assumption, we have $K^{(0)}_{\alpha\beta}=0$ which entails, via (\ref{eq:electric-magnetic}), 
$B^{(0)}_{\alpha\beta}=0$. Hence from (\ref{eq:scalarP}) we get $P|_{\Sigma}=0$ and thus
\begin{equation}
\nabla_a\rho\nabla^a\rho|_{\Sigma}\geq 0.
\label{eq:drho-sigma}
\end{equation}
We can now compute the explicit value of the scalar $\nabla_a\rho\nabla^a\rho$ at any point of the Schwarzschild space-time if we choose a coordinate representation. For example, we may choose the standard Schwarzschild coordinates
\begin{equation}
ds^2=-\left(1-\frac{2M}{r}\right)dt^2+\frac{dr^2}{\left(1-\frac{2M}{r}\right)} 
+r^2(d\theta^2+\sin^2\theta d\varphi^2)\;,\quad
0<\theta<\pi\;,\quad 0<\varphi<2\pi,
\label{eq:sch-coordinates}
\end{equation}
where $M$ is the Schwarzschild mass. As is very well-known
when $r>2M$ the Schwarzschild coordinates cover either 
of the two asymptotically flat exterior regions ${\mathcal D}_1$ and ${\mathcal D}_2$  whereas for $0<r<2M$ the coordinates cover either of the two interior regions ${\mathcal B}^+$ or 
${\mathcal B}^-$. In the Schwarzschild coordinates one has
\begin{equation}
\nabla_a\rho\nabla^a\rho=-\frac{9M^2(2M-r)}{r^9}. 
\label{eq:value-of-drho}
\end{equation}
Note that since the scalar $\nabla_a\rho\nabla^a\rho$ is an invariant quantity,
the relation (\ref{eq:value-of-drho}) gives its value at any point of the maximal extension $\mathcal M$, including the region not covered by the Schwarzschild coordinates 
(this region is comprised by the points such that $r=2M$ which correspond to the region ${\mathcal H}^+\cup{\mathcal H}^-$). Now, from (\ref{eq:value-of-drho}) it is straightforward to see
that in the interior of any of the black hole regions the inequality $\nabla_a\rho\nabla^a\rho<0$ holds and combining this property with (\ref{eq:drho-sigma}) the required result follows.\qed  
\begin{corollary}
Any inextensible connected time-symmetric $C^k$, $k>2$ hypersurface $\Sigma$ embedded in the Schwarz\-schild space-time must intersect the bifurcation sphere.
\label{cor:hypersurface-no-boundary}  
\end{corollary}
\proof
Proposition \ref{prop:time-symmetric} implies that $\Sigma$ must be included
in the complementary set of the interior of the black hole regions, which is the set 
$D_1\cup D_2\cup{\mathcal H}^+\cup{\mathcal H}^-$. 
Moreover since $\Sigma$ is space-like and inextensible, the intersection of 
$\Sigma$ and the set $({\mathcal H}^+\cup{\mathcal H}^-)\backslash\Omega$ must be the empty set from which we conclude that the intersection of $\Sigma$ and $\Omega$ is never empty. \qed

\begin{corollary}
A point in the event horizon, but not in the bifurcation sphere, 
never belongs to the boundary of a connected, time-symmetric, analytic hypersurface $\Sigma$ embedded in any of the exterior regions of the Schwarzschild space-time. 
\end{corollary}
\proof Suppose the contrary and let $\Sigma\cap({\mathcal H}\backslash\Omega)$ be non-empty, where
${\mathcal H}\equiv{\mathcal H}^+\cup{\mathcal H}^-$. Since $\Sigma$ is analytic, we may construct an analytic extension $\tilde{\Sigma}$ of $\Sigma$ through the set 
$\Sigma\cap({\mathcal H}\backslash\Omega)$ such that $\Sigma\subset\tilde{\Sigma}$ and
$\tilde{\Sigma}\cap\mbox{int}({\mathcal B})\neq\varnothing$, where 
${\mathcal B}\equiv{\mathcal B}^+\cup{\mathcal B}^-$. We should distinguish
now two possibilities: either $\tilde{\Sigma}$ is space-like at every point or it contains a subset in which it is null. In the first case one can define a symmetric tensor 
field $K_{\alpha\beta}$ on $\tilde{\Sigma}$ through Eq. (\ref{eq:first-second-form}) which 
by construction is analytic and vanishes in a proper open subset of $\tilde{\Sigma}$ which is impossible. In the second alternative we can still use Eq. (\ref{eq:first-second-form}) to define an analytic symmetric tensor field $\tilde{K}_{\alpha\beta}$ at any point of $\tilde{\Sigma}$ if we remove the normalisation condition on the vector $\vec{\boldsymbol N}$. On those points belonging to $\Sigma$ is easy to check that the relation 
$$
\tilde{K}_{\alpha\beta}=\Psi K_{\alpha\beta}\;,
$$
holds, where $K_{\alpha\beta}$ is the tensor obtained with the normalisation condition on $\vec{\boldsymbol N}$ and $\Psi$ is a certain scalar function. Since $K_{\alpha\beta}=0$ we get $\tilde{K}_{\alpha\beta}=0$ on $\Sigma$, which again is impossible for an analytic tensor on $\tilde{\Sigma}$.
\qed 

\section{Applications and conclusions}
\label{sec:conclusions}
In this work we have shown that inextensible time-symmetric hypersurfaces embedded in the Schwarzschild-Kruskal space-time never intersect the black hole region and always intersect the bifurcation sphere. An interesting issue not solved in this work (nor, as far as we know, in any other place of the literature) deals with the uniqueness of embedded time-symmetric hypersurfaces in the Schwarzschild-Kruskal space-time. It is tempting to conjecture that the integral hypersurfaces of the static Killing vector are the only time-symmetric hypersurfaces and one expects that our techniques should contribute to settle this question. If this conjecture were true then one could seek an uniqueness result of the Schwarzschild space-time formulated in terms of the uniqueness in the existence of embedded time-symmetric hypersurfaces. 

As mentioned in section \ref{sec:preliminaries}  a time-symmetric hypersurface has been regarded as space-like by definition. However, the second fundamental form can be defined for any kind of hypersurface 
by means of Eq. (\ref{eq:first-second-form}) (the vector
$\vec{\boldsymbol N}$ need not be normalised) and therefore a logical question would be what would happen when arbitrary 
embedded hypersurfaces with vanishing second fundamental form are considered
(this condition is independent from the normalisation for 
$\vec{\boldsymbol N}$ chosen). 
When we deal with a time-like hypersurface, then equations 
(\ref{eq:decompose-rho}) and (\ref{eq:scalarP})-(\ref{eq:vectorP}) adopt a similar form and the quantity $P$ does vanish if the second fundamental form is zero but $P_a$ could still be time-like or space-like. Thus the sign of the right hand side of (\ref{eq:grad-rho-norm}) is not determined in general and hence the main argument used in the proof of proposition \ref{prop:time-symmetric} is no longer valid. Therefore we would have to consider the possibility in which the hypersurface has null regions and space-like regions. In the null regions one does not have a well-defined orthogonal splitting with respect to the null vector orthogonal to the hypersurface and hence it is not possible to decompose 
$\nabla_a\rho$ on these regions in the manner shown in eq. 
(\ref{eq:decompose-rho}).

\section{Acknowledgments}
We thank Dr. Juan Antonio Valiente Kroon for a careful reading of the manuscript and useful comments and we also thank an anonymous referee for constructive criticism. 
We acknowledge the financial support provided by the Research Centre of  
Mathematics of the University of Minho (Portugal) through the ``Funda\c{c}\~ao para a Ci\^encia e a Tecnologia'' (FCT) Pluriannual  
Funding Program. We also thank the Albert Einstein Institut f\"ur Gravitationsphysik in Golm (Germany), where part of this work was carried out, for hospitality and financial support. 

\bibliographystyle{amsplain}

\bibliography{/home/alfonso/trabajos/BibDataBase/Bibliography}

\end{document}